# Ovarian cancer dissemination reveals a convergent immune ecosystem

Yangyang Zhang[1,2,†], Zilong Wang[3,4,†], Stéphanie Nougaret[5,6,†], Zhihao Zhang[2,7], Yongqi Chen[1,2], Chu Han[8,9], Ya Zhang[10], Xiang Zhang[11], Xuxia Zhang[12], Xingfa Chen[13], Xiaochun Wan[2,14,15], Zijing Lin[2,7,16], Shuai Liu[2,17], Hui Li[2,14], Jun Zhu[1,2], Zheng Feng[1,2], Hao Wen[1,2], Yanping Zhong[1,2], Yunqiu Zhang[18], Jie Duan[19], Jun Li[19], Haiyan Zhu[18], Wei Jiang[19], Xingzhu Ju[1,2], Xiaohua Wu[1,2,*], Shuo Wang[20,21,*], Haiming Li[2,7,*], Qinhao Guo[1,2,*]

[1]Department of Gynecologic Oncology, Fudan University Shanghai Cancer Center, Fudan University, Shanghai 200032, China.

[2]Department of Oncology, Shanghai Medical College, Fudan University, Shanghai 200032, China.

[3]Microsoft Research Asia, Shanghai 200232, China.

[4]CataX AI, 900 Yishan Road, Shanghai 200233, China

[5]Department of Radiology, Montpellier Cancer Institute, University of Montpellier, Montpellier 34090, France.

[6]PINKcc Lab, INSERM U1194, Montpellier Cancer Research Institute, University of Montpellier, Montpellier, France.

[7]Department of Radiology, Fudan University Shanghai Cancer Center, Fudan University, Shanghai 200032, China.

[8]Medical Research Institute, Guangdong Provincial People's Hospital (Guangdong Academy of Medical Sciences), Southern Medical University, Guangzhou 510080, China.

[9]Department of Radiology, Guangdong Provincial People's Hospital (Guangdong Academy of Medical Sciences), Southern Medical University, Guangzhou 510080, China.

[10]Department of Radiology, The Third Affiliated Hospital of Kunming Medical University, Yunnan Cancer Hospital, Kunming 650118, China.

[11]Department of Radiology, Sun Yat-sen Memorial Hospital, Sun Yat-sen University, Guangzhou 510120, China.

[12]Medical Imaging Center, Gansu Provincial Maternity and Child-care Hospital (Gansu Provincial Central Hospital), Lanzhou 730050, China.

[13]Department of Radiology, Clinical Oncology School of Fujian Medical University, Fujian Cancer Hospital, Fuzhou 350014, China.

[14]Department of Pathology, Fudan University Shanghai Cancer Center, Fudan University, Shanghai 200032, China.

[15]Institute of Pathology, Fudan University Shanghai Cancer Center, Fudan University, Shanghai 200032, China.

[16]Department of Radiology, Jinshan Hospital, Fudan University, Shanghai 201508, China.

[17]Department of Nuclear Medicine, Fudan University Shanghai Cancer Center, Fudan University, Shanghai 200032, China.

[18]Department of Gynecology, Shanghai Key Laboratory of Maternal-Fetal Medicine, Shanghai Institute of Maternal-Fetal Medicine and Gynecologic Oncology, Shanghai First Maternity and Infant Hospital, School of Medicine, Tongji University, Shanghai 200092, China.

[19]Institute of Obstetrics and Gynecology, Hospital of Obstetrics and Gynecology, Fudan University, Shanghai 200080, China.

[20]Digital Medical Research Centre, School of Basic Medical Sciences, Fudan University, Shanghai 200032, China.

[21]Shanghai Key Laboratory of MICCAI, Shanghai 200032, China.

[†]These authors contributed equally.

*Corresponding authors.

E-mail addresses: wu.xh@fudan.edu.cn (Xiaohua Wu), shuowang@fudan.edu.cn (Shuo Wang), lihaiming0109@163.com (Haiming Li), guoqinhao_sh@fudan.edu.cn (Qinhao Guo).

**Abstract**

Advanced ovarian cancer disseminates extensively across diverse anatomical sites, forming lesions with varied morphologies. However, the organisation of this dissemination remains poorly understood. Here we systematically mapped metastatic sites and lesion morphologies in 5,587 patients, identifying a miliary–diffuse dissemination pattern (MDP) associated with shorter survival even after complete macroscopic resection (adjusted overall-survival hazard ratio, 1.96). Molecular profiling revealed that MDP-high primary tumours exhibited attenuated interferon-response and HLA antigen-presentation programmes despite heterogeneous genomic backgrounds. Inferred HLA-I allele-copy loss in whole-genome doubling (WGD)-negative tumours and WGD-associated epigenetic alterations suggest potentially distinct routes to this shared immune phenotype. Single-cell and spatial analyses localised this attenuation to malignant cells within a tumour microenvironment characterised by vascular–stromal remodelling and reduced intratumoural CD8+ T-cell infiltration. Although defined by multisite dissemination, MDP could be predicted preoperatively from primary-tumour MRI, with external validation across four centres (pooled AUC, 0.814). Together, these findings link the macroscopic architecture of ovarian cancer dissemination to convergent immune attenuation in primary tumours, with prognostic significance beyond complete resection.

## Introduction

Advanced ovarian cancer is characterised by extensive peritoneal dissemination and poor prognosis. Despite cytoreductive surgery and platinum-based chemotherapy, disease recurrence remains common.[1-3] The distribution and morphology of tumour deposits influence surgical planning and the likelihood of complete resection.[4] Nevertheless, baseline disease burden remains associated with survival even after complete macroscopic resection.[5] These observations suggest that the anatomical and morphological organisation of tumour spread may reflect clinically meaningful tumour aggressiveness beyond surgical resectability.

Our previous imaging studies distinguished mass-like and infiltrative morphologies of primary high-grade serous ovarian cancer (HGSOC) and established their value in predicting surgical resectability.[6,7] We further demonstrated the potential of MRI radiomics to predict radiologically occult miliary changes in the small-bowel mesentery,[8] and extended morphological assessment across metastatic sites to improve resectability prediction and reveal transcriptomic differences between primary-tumour morphologies.[9] Beyond imaging, clinical and pathological studies have linked miliary spread and anatomical dissemination patterns to clinical outcomes and molecular subtypes.[10-13] Most previous studies, however, have examined individual features or predefined categories of spread. Whether large-scale, patient-level mapping can reveal recurrent patterns of coexisting dissemination features with prognostic significance beyond surgical resectability remains unknown.

Multiregion sequencing has revealed divergent routes of clonal spread and intraperitoneal mixing in HGSOC, while single-cell and spatial studies have uncovered site-specific immune evasion and interactions among malignant, immune and stromal cells.[14-18] Structural genomic evolution is also heterogeneous,[19] with whole-genome doubling (WGD) linked to STING1 repression and attenuated interferon signalling.[20] Yet these genomic and cellular insights remain largely disconnected from patient-level patterns of tumour dissemination. This knowledge gap has hindered the development of a systematic framework for interpreting dissemination patterns in relation to tumour biology, clinical risk and patient management.

In this study, we combined AI-assisted record extraction with data-driven analysis to map tumour dissemination in 5,587 patients with advanced epithelial ovarian cancer, identifying four recurrent, coexisting patterns from multimodal clinical data. Among these, the miliary–diffuse dissemination pattern (MDP) was associated with platinum resistance, lower complete-resection rates and shorter survival even after complete macroscopic resection. We then integrated multi-omic, single-cell and spatial analyses to characterise this high-risk pattern, revealing WGD enrichment, shared attenuation of malignant-cell interferon-response and HLA-I programmes across genomic backgrounds, and an associated vascular–stromal ecosystem with reduced intratumoural CD8+ T-cell infiltration. Finally, primary-tumour MRI radiomics predicted this multisite dissemination phenotype preoperatively and stratified survival in a pooled four-centre external cohort.

## Results

### A multimodal atlas of ovarian cancer dissemination

We constructed a patient-level atlas of ovarian cancer dissemination using records from 5,587 patients with FIGO stage III–IV epithelial ovarian cancer treated at Fudan University Shanghai Cancer Center (FUSCC) (**Fig. 1a**, **Extended Data Fig. 1** and **Supplementary Table 1**). Using a human-in-the-loop workflow, we integrated imaging, operative and pathological records (**Supplementary Table 2**) across 66 anatomical sites and nine non-exclusive morphological categories (**Fig. 1b** and **Supplementary Tables 3 and 4**). Inter-radiologist agreement on site involvement was high (pre-adjudication Cohen's $\kappa = 0.999$ across 368,742 patient–site pairs), and blinded review of source images and histopathology in 360 randomly selected patients showed complete concordance with the original reports for the assessed findings (**Supplementary Tables 5–7**). The resulting 182-feature matrix summarised each patient's anatomical involvement and lesion morphology (**Methods**).

The Douglas-pouch peritoneum and greater omentum were among the most frequently involved sites (89.4% and 82.7% of patients, respectively). Perihepatic liver-capsule involvement was recorded in 33.4% of patients, whereas hepatic and pulmonary parenchymal involvement occurred in 6.3% and 1.3%, respectively (**Supplementary Table 3**). Within the peritoneal, mesenteric and gastrointestinal module, nodular, diffuse and miliary spread were recorded in 90.2%, 46.3% and 33.3% of patients, respectively (**Fig. 1c**).

Regional anatomical involvement was associated with clinical outcomes across hepatobiliary, splenic, lymph-node and other abdominopelvic and distant sites (**Fig. 1d**). In exploratory univariable analyses, all four anatomical categories examined were associated with lower odds of complete resection (R0), higher odds of primary platinum resistance and shorter progression-free survival (PFS). For example, hepatobiliary involvement was associated with lower odds of R0 (odds ratio (OR), 0.56; 95% confidence interval (CI), 0.49–0.64) and shorter PFS (hazard ratio (HR), 1.26; 95% CI, 1.17–1.37). Associations with overall survival (OS) were less uniform, with splenic and nodal involvement not remaining significant after multiple-testing correction. Associations of hepatobiliary involvement with lower odds of R0 and shorter PFS persisted in multivariable models (adjusted OR, 0.56; 95% CI, 0.49–0.64; adjusted HR, 1.20; 95% CI, 1.11–1.31; **Supplementary Table 8**). Among molecularly annotated patients, four dissemination features were less frequent in homologous-recombination-deficient (HRD) than homologous-recombination-proficient (HRP) tumours, including plaque-like pleural lesions or surface scalloping, diffuse greater-omental spread, and involvement of the subdiaphragmatic splenic and perihepatic liver capsules (all $q < 0.05$; **Fig. 1e**).

Individual sites and site–morphology combinations showed distinct associations with clinical outcomes (**Fig. 1f–i**). Notably, miliary spread on the small-bowel serosa was associated with markedly lower odds of R0 (OR, 0.22; 95% CI, 0.17–0.28) and shorter OS (HR, 1.76; 95% CI, 1.42–2.17). Porta-hepatis lymph-node involvement was associated with primary platinum resistance (OR, 2.05; 95% CI, 1.53–2.76), whereas greater-omental involvement was associated with shorter PFS and OS. These associations remained significant after multiple-testing correction. Although individual dissemination features provided clinically relevant information, whether they formed recurrent, organised patterns across patients remained unclear.

## Four coexisting patterns of tumour spread

We applied repeated non-negative matrix factorisation (NMF) to the 182-feature matrix to identify recurrent combinations of anatomical involvement and morphology (**Fig. 2a**). Components recurring across runs and factorisation ranks ranging from 4 to 15 were grouped by overlap among their highest-weighted features, producing four consensus dissemination patterns (MP1–MP4) independently of clinical outcomes (**Fig. 2b, Supplementary Fig. 2a,b, Supplementary Table 9** and **Methods**).

The four patterns captured distinct anatomical and morphological configurations (**Fig. 2b–d**). MP1 (lymphatic–nodal) combined nodal involvement across pelvic, retroperitoneal and extra-abdominal stations, whereas MP2 (pelvic–omental) coupled pelvic involvement with nodular peritoneal and omental spread. MP3 (upper-abdominal) emphasised hepatobiliary, diaphragmatic and splenic involvement. MP4, termed the **miliary–diffuse dissemination pattern (MDP)**, combined miliary and diffuse involvement of the mesentery and small-bowel serosa as well as other peritoneal surfaces. For each pattern, the feature co-occurrence network was denser than networks derived from size-matched random feature sets (**Supplementary Fig. 2e**).

We next quantified the representation of each pattern in individual patients by jointly projecting their dissemination profiles onto the four consensus patterns using non-negative least squares. Each patient thus received four continuous pattern scores, allowing multiple patterns to coexist within individuals. Across the six pairs of patterns, 5.1–7.5% of patients had scores above the respective upper-quartile thresholds for both patterns (**Supplementary Table 10**). Resampling the constituent NMF components largely preserved patient-score rankings and pattern composition (**Supplementary Fig. 2c,d**). We further assessed pattern robustness by refitting NMF under histological restriction and evidence-state masking (**Supplementary Fig. 3a**). Separately, when patient scores were recalculated using the original patterns after masking uncertain, unassessed or both categories of findings, MDP-score rankings in HGSOC remained highly correlated with the original rankings (Spearman $\rho = 0.991$, 0.960 and 0.942, respectively; **Supplementary Fig. 3c**).

The four patterns showed distinct clinical associations in the full cohort of 5,587 patients with advanced epithelial ovarian cancer (**Fig. 2e**). After correction across 16 pattern–endpoint comparisons, all four were associated with lower odds of R0 resection and shorter PFS in univariable analyses. Only MP2 and MDP were significantly associated with OS. MDP was associated with adverse outcomes across all four endpoints, including primary platinum resistance among the 3,510 evaluable patients. In univariable models, each standard-deviation increase in MDP score was associated with lower odds of R0 (OR, 0.54; 95% CI, 0.51–0.58) and shorter OS (HR, 1.25; 95% CI, 1.19–1.32). MDP remained associated with all four clinical endpoints in joint models including the other three pattern scores and clinical covariates (**Supplementary Fig. 2f–i**). Together, these findings identified MDP as a dissemination pattern associated with high clinical risk, motivating its focused investigation in HGSOC.

**MDP-associated survival disadvantage persists after complete resection**

We next investigated whether MDP captured prognostic risk in HGSOC beyond its association with surgical resectability. Using a Gaussian mixture model fitted to MDP scores in all 5,080 patients with HGSOC, we classified 2,085 patients as MDP-high and 2,995 as MDP-low (**Fig. 3a** and **Supplementary Table 11**). The two-component model showed high classification stability across bootstrap resamples (median out-of-bag adjusted Rand index, 0.963; **Supplementary Fig. 4**). MDP-high patients more frequently exhibited miliary spread across the mesentery and peritoneal surfaces, together with diffuse involvement of the small-bowel serosa and other gastrointestinal sites (**Fig. 3b**).

Complete macroscopic resection (R0) was achieved in 65.9% of MDP-high patients, compared with 84.5% of MDP-low patients (**Fig. 3c**). MDP-high status remained associated with lower odds of R0 after adjustment for age, FIGO stage and surgical pathway (OR, 0.36; 95% CI, 0.31–0.41), with consistent associations in patients undergoing primary debulking surgery (PDS) and interval debulking surgery (IDS). MDP-high patients also had shorter PFS and OS (**Fig. 3d**), with adjusted HRs of 1.46 (95% CI, 1.34–1.59) and 1.79 (95% CI, 1.58–2.03), respectively, after adjustment for age, FIGO stage, surgical pathway and R0 status (**Extended Data Fig. 2a,b**). These survival associations were consistent across both surgical pathways (**Supplementary Table 12**).

Notably, the association with shorter survival persisted among the 3,906 patients who achieved complete macroscopic resection. Within this group, MDP-high patients had shorter PFS (HR, 1.43; 95% CI, 1.29–1.58) and OS (HR, 1.78; 95% CI, 1.53–2.07; **Fig. 3e**). These associations remained after adjustment for age, FIGO stage and surgical pathway, with adjusted HRs of 1.56 (95% CI, 1.41–1.73) for PFS and 1.96 (95% CI, 1.68–2.28) for OS (**Supplementary Table 13**). MDP remained associated with clinical outcomes when analysed as a continuous score and after additional adjustment for recorded miliary spread, diffuse spread and mesenteric involvement (**Supplementary Table 14**).

MDP status was also associated with differences in preoperative circulating markers. Among 2,979 patients treated with PDS, MDP-high patients had higher CA125 and HE4 levels, neutrophil-to-lymphocyte ratios and platelet counts, together with lower albumin levels and lymphocyte counts (**Extended Data Fig. 2c**). Plasma cell-free RNA profiling in 157 patients revealed enrichment of a platelet-activation programme in MDP-high patients, whereas MYC-target and G2M-checkpoint programmes were enriched in MDP-low patients, with all three programmes remaining significant after multiple-testing correction (**Extended Data Fig. 2d**). These clinical and circulating molecular findings prompted further investigation of the tumour biology associated with MDP.

## Convergent immune attenuation across genomic backgrounds

To investigate the tumour biology associated with MDP-related risk, we analysed whole-genome sequencing profiles from 135 tumour specimens representing 134 patients (**Fig. 4a** and **Supplementary Tables 15 and 16**). Whole-genome doubling (WGD) was considerably more frequent in MDP-high than MDP-low patients (42/58, 72.4% versus 18/76, 23.7%; $P = 2.79 \times 10^{-8}$; **Fig. 4b**). This association was observed in both homologous-recombination-deficient (HRD) and homologous-recombination-proficient (HRP) backgrounds. Within HRD, 68% of WGD-positive patients were MDP-high, compared with 29% of WGD-negative patients; the corresponding proportions within HRP were 74% and 4%.

To identify molecular programmes associated with MDP, we integrated RNA, protein and metabolite measurements from 82 FUSCC tumours obtained at primary debulking surgery (PDS) using MOFA+ (**Fig. 4c**). Neither dissemination labels nor WGD status were included as model inputs. Seven retained axes described recurrent multi-omic variation (**Extended Data Fig. 3a–c**). Higher stromal-remodelling scores were associated with greater odds of MDP-high status (OR per standard-deviation increase, 2.54; 95% CI 1.51–4.26; $q = 0.003$), whereas higher antigen-presentation scores were associated with lower odds (OR 0.37; 95% CI 0.19–0.72; $q = 0.012$; **Fig. 4d**). The stromal axis linked fibroblast-associated extracellular-matrix programmes with TGF-β-related activity, whereas the antigen-presentation axis encompassed interferon responses, HLA expression and antigen-processing programmes (**Fig. 4e** and **Extended Data Fig. 3d**). Both associations persisted in a joint model accounting for tumour purity, HRD and WGD status (adjusted OR, 4.04; 95% CI, 1.51–10.77 for stromal remodelling; adjusted OR, 0.18; 95% CI, 0.06–0.56 for antigen presentation; **Supplementary Table 17**).

In keeping with the antigen-presentation axis, MDP-high tumours showed lower HLA-I and HLA-II complex scores and interferon-γ (IFN-γ)-response scores at both RNA and protein levels (**Fig. 4f**). These analyses included 82 tumours with RNA data and 78 with protein data, and all six associations remained significant after correction across the nine displayed endpoints ($q < 0.003$).

Notably, MDP-high tumours showed lower HLA-I and IFN-γ-response scores than MDP-low tumours, including among WGD-positive specimens without detectable HLA-I allele-copy loss at assessable loci (29 evaluable for RNA; 27 for protein; **Fig. 4g**). HLA-II scores showed the same direction, although the RNA-level difference did not reach significance ($P = 0.085$). Among WGD-negative specimens, inferred HLA-I allele-copy loss was more frequent in MDP-high than MDP-low tumours (5/7 versus 5/35, OR 13.58, exact 95% CI 1.69–180.01, $P = 0.0049$, **Fig. 4h**). Clonal reconstruction further showed that 6p21/MHC-region loss was not detected in the first WGD clone in 21 of 28 evaluable WGD events (75%; **Extended Data Fig. 4a,b**).

We further examined WGD-associated cellular and regulatory changes using complementary published experimental and epigenetic datasets. Patients with expanded WGD clones showed higher micronucleus frequencies than those without expanded WGD clones ($P = 4.1 \times 10^{-4}$; **Extended Data Fig. 4c**). Longitudinal profiling of mouse mammary tumours identified WGD-associated differences in IFN-γ–JAK–STAT1 signalling, NLRC5/CIITA and MHC-I/MHC-II antigen-presentation programmes, while single-cell ATAC-seq showed altered gene activity at selected antigen-presentation loci (**Extended Data Fig. 4d,e**). At the epigenetic level, hypermethylated regions were enriched for interferon-response and immune-related pathways (**Extended Data Fig. 4f**). WGD-positive tumours also showed higher STING1 and GBP4 promoter methylation in two independent ovarian cancer cohorts (TCGA-OV, $n = 386$; GSE292140-OV, $n = 154$; **Extended Data Fig. 4g**).

Together, these findings showed that MDP was associated with WGD enrichment, stromal remodelling and a shared attenuation of interferon-response and antigen-presentation programmes across heterogeneous genomic backgrounds. The persistence of immune attenuation without detectable HLA-I copy loss, alongside candidate HLA-I losses and WGD-associated epigenetic differences, points to distinct genomic and regulatory contexts for this phenotype. We next investigated the cellular states and spatial organisation underlying this phenotype.

**Malignant-cell immune attenuation and vascular–stromal organisation in MDP**

To identify the cellular location of MDP-associated immune attenuation and examine its tissue context, we analysed an integrated single-cell atlas comprising 2,871,618 cells from 682 samples across 30 datasets (**Supplementary Tables 18–20**). Benchmarking favoured scVI for balancing batch correction and biological conservation (**Extended Data Fig. 5a,b**). Single-cell profiles with matched dissemination annotations were available from treatment-naive primary tumours of 57 FUSCC patients, including 21 MDP-high and 36 MDP-low patients (**Methods**). Malignant cells from MDP-high tumours showed lower interferon-response and major histocompatibility complex (MHC) class I and II programme scores in patient-level comparisons, supported by concordant pseudobulk analyses (**Fig. 5a**; gene sets in **Supplementary Table 21**). All 17 examined HLA genes were expressed at lower levels in MDP-high malignant cells after multiple-testing correction within the HLA panel (**Fig. 5b**). Inferred STAT–IRF and RFX5 activities were also lower, alongside reduced response scores for IFN-γ, type I interferons and IL-27 (**Fig. 5c,d** and **Supplementary Tables 22 and 23**).

We also examined ascites-associated clinical features and tumour-cell programmes to obtain complementary evidence. In the PDS clinical subset, higher MDP scores were associated with greater ascites prevalence and volume and a higher frequency of positive cytology (**Extended Data Fig. 6a–c**). An external tumour-cell transcriptomic cohort showed lower IFN-γ and antigen-presentation programmes in samples annotated as having miliary spread, both overall and within ascites-derived samples (**Extended Data Fig. 6d**). In nine patients with paired ascites fractions, naturally formed tumour-cell spheroids showed higher adhesion and extracellular-matrix programmes but lower immune-recognition programmes than single-cell fractions (**Extended Data Fig. 6e**). Cytological images further illustrated malignant-cell aggregates in ascites (**Extended Data Fig. 6f**).

Spatial profiling provided further evidence of immune attenuation in tumour-enriched regions. Across 34 pan-cytokeratin-enriched areas of interest from 10 patients, regions from MDP-high tumours showed lower IFN-γ, MHC-I/II and antigen-processing programme scores (**Fig. 5e**). Patient-level comparisons showed concordant reductions across all five programmes, including IFN-γ response (**nominal** $P = 0.048$; **Supplementary Table 24**).

We next explored associations between malignant-cell immune attenuation and non-malignant cell-state organisation. Using lymphoid, myeloid, fibroblast and endothelial states annotated in the integrated atlas (**Extended Data Figs. 7 and 8** and **Supplementary Table 19**), we analysed 108 treatment-naive primary HGSOC patient profiles (61 institutional and 47 public). Patient-resampling and consensus analyses resolved six tumour-microenvironment ecosystems (**Extended Data Fig. 9a,b**). Among these, TME-Angio was characterised by tip endothelial cells, pericytes and fibroblasts (**Fig. 5f** and **Supplementary Tables 19 and 25**).

Among 51 patients with matched MDP labels, exploratory analysis suggested higher TME-Angio activity in MDP-high tumours (nominal $P = 0.037$, $q = 0.223$; **Fig. 5g** and **Extended Data Fig. 9c**). Malignant-cell MHC-I scores differed across the six ecosystems ($P = 0.0022$; **Extended Data Fig. 9d**), and TME-Angio activity was inversely correlated with all five examined interferon and antigen-presentation programmes (**Extended Data Fig. 9e**). The inverse association with MHC-I persisted after adjustment for MDP group (partial Spearman $\rho = -0.50$; permutation $P = 0.0002$; **Fig. 5h**).

We then examined the spatial organisation of TME-Angio in eight external spatial-transcriptomic sections obtained after neoadjuvant chemotherapy, without MDP annotations. In these sections, TME-Angio was associated with myeloid-enriched regions and lower relative CD4+ and CD8+ T-cell abundance (**Fig. 5i**). These inverse associations with T-cell abundance were consistent across spot-level, neighbourhood and proximity analyses. In MDP-annotated spatial profiles, SMA-enriched regions showed higher collagen-crosslinking and lower chemokine-response scores, whereas CD3-enriched regions showed altered chemotaxis and NK-response programmes (**Fig. 5j** and **Supplementary Table 24**).

Histological and multiplex imaging analyses provided complementary assessments of lymphocyte abundance and tumour–stroma compartmentalisation in MDP-annotated tumours. Among 42 patients, MDP-high tumours had lower tumour-associated lymphocyte density and a higher stromal-to-epithelial lymphocyte-density ratio ($P = 0.019$ and 0.001, respectively; **Fig. 5k** and **Supplementary Table 26**). Multiplex imaging in 63 patients showed lower tumour-compartment fractions of CD8+, CD4+, CD20+ and CD4+FOXP3+ cells, together with a reduced tumour-to-stroma CD8+ ratio (**Fig. 5l** and **Supplementary Table 26**). The proportion of regulatory T cells among CD4+ cells did not differ significantly between groups ($P = 0.653$). Together, these findings localise MDP-associated immune attenuation to malignant cells and link it to a vascular–stromal ecosystem characterised by reduced intratumoural lymphocyte infiltration. We next investigated whether this clinically aggressive dissemination phenotype could be recognised before surgery.

## Primary-tumour MRI predicts MDP and stratifies survival

Given that MDP-associated molecular and cellular features were identified in primary tumours, we investigated whether primary-tumour MRI could predict this multisite dissemination phenotype, which may not be fully apparent on conventional preoperative imaging. The imaging cohort comprised 282 FUSCC patients and 261 patients from four external centres, all with FIGO stage III–IV HGSOC treated with primary debulking surgery (PDS; **Fig. 6a**, **Supplementary Fig. 5** and **Supplementary Table 27**). We developed an XGBoost classifier using shape, intensity and texture features extracted from diffusion-weighted MRI of primary tumours acquired across different scanner platforms and imaging protocols (**Supplementary Table 28**). Nested cross-validation with five outer folds yielded an area under the receiver-operating-characteristic curve (AUC) of 0.814 (95% CI 0.764–0.861; **Fig. 6b**), compared with 0.775 for elastic-net logistic regression and 0.762 for linear support-vector machines evaluated on the same folds. Feature-attribution analysis identified texture and shape measurements, including size-zone non-uniformity and sphericity (**Extended Data Fig. 10a**).

To assess generalisability, we applied the frozen XGBoost model without refitting to four geographically distinct external cohorts: Yunnan Cancer Hospital (n = 102), Fujian Cancer Hospital (n = 37), Gansu Provincial Maternity and Child Care Hospital (n = 50) and Sun Yat-sen Memorial Hospital (n = 72). The model distinguished MDP-high from MDP-low patients with a pooled external AUC of 0.814 (95% CI 0.755–0.868), with centre-specific AUCs ranging from 0.732 to 0.826 (**Fig. 6c**). At the locked development-cohort threshold of 0.6088, the model achieved a sensitivity of 75.0% and specificity of 80.6% in the pooled external cohort, with positive and negative predictive values of 69.2% and 84.7%, respectively (Supplementary Table 29). Calibration analysis indicated overall overestimation of the probability of MDP-high status across external centres (Supplementary Fig. 6). These results suggest that imaging features of the primary tumour can discriminate a multisite dissemination phenotype across independent centres.

We next examined whether MRI-predicted MDP also stratified survival. Using the same locked threshold, 108 of 282 internal patients and 104 of 261 external patients were classified as MRI-predicted MDP-high. In the internal cohort, MRI-predicted MDP-high status was associated with shorter progression-free survival (PFS; HR 1.63, 95% CI 1.07–2.49) and overall survival (OS; HR 2.98, 95% CI 1.34–6.65; **Fig. 6d**). Similar associations were observed in the pooled external cohort, with HRs of 1.68 (95% CI 1.17–2.39) for PFS and 2.22 (95% CI 1.43–3.47) for OS (**Fig. 6e**).

The prognostic associations remained after adjustment for age, FIGO stage, R0 status, CA125 and HE4 (**Extended Data Fig. 10b,c**). External models additionally stratified baseline hazards by centre, yielding adjusted HRs of 1.55 for PFS (95% CI 1.07–2.25) and 2.16 for OS (95% CI 1.35–3.44). Centre-specific estimates pointed in the same direction in all four centres, with unadjusted OS hazard ratios of 2.11–2.66 (**Extended Data Fig. 10d,e**). Together, these findings suggest that primary-tumour MRI can predict a patient-level dissemination phenotype before surgery and stratify subsequent survival risk across independent cohorts.

## Discussion

Understanding how metastatic disease is organised across anatomical sites, and what this organisation reveals about underlying tumour biology, remains a fundamental challenge in cancer research.[21-23] Here we present a systematic atlas of ovarian cancer dissemination in 5,587 patients, identifying four recurrent patterns that can coexist within individuals. By capturing how metastatic lesions are distributed and organised across anatomical sites, this atlas extends beyond conventional descriptions of tumour spread. Among these patterns, the miliary–diffuse dissemination pattern (MDP) was associated with adverse survival even after complete macroscopic resection. MDP-high tumours showed marked enrichment of whole-genome doubling and convergent attenuation of malignant-cell immune programmes across heterogeneous genomic backgrounds, with genetic and epigenetic findings suggesting distinct routes to this shared immune state. Single-cell and spatial analyses further linked MDP to a vascular–stromal ecosystem with reduced intratumoural lymphocyte infiltration, while primary-tumour MRI enabled preoperative prediction of MDP. Together, these findings support dissemination architecture as a biologically informative dimension of ovarian cancer, connecting macroscopic metastatic patterns with tumour biology and clinical behaviour.

Previous studies of ovarian cancer dissemination have largely examined predefined anatomical regions or individual lesion morphologies, often considering imaging, operative and pathological findings separately.[6,8,9,12,24-26] Our integrated atlas identifies four recurrent, coexisting patterns with recognisable clinical and biological correlates. The lymphatic–nodal pattern (MP1) recalls the more favourable prognosis reported for node-limited disease compared with extensive peritoneal dissemination,[27] whereas the pelvic–omental pattern (MP2) involves a metastatic niche in which adipocytes and immune cells facilitate tumour colonisation.[28,29] The upper-abdominal pattern (MP3) encompasses sites that complicate complete cytoreduction, while MDP captures a recurrent patient-level pattern of miliary and diffuse involvement of the mesentery and small-bowel serosa, together with spread across other peritoneal surfaces.[10,30,31] Notably, although all four patterns were associated with lower complete-resection rates and shorter progression-free survival, only MP2 and MDP showed significant associations with overall survival. At the individual-feature level, diffuse omental and perihepatic involvement were also less frequent in homologous-recombination-deficient than -proficient tumours, suggesting a relationship between dissemination characteristics and genomic background.[7,32,33] These observations underscore the clinical and biological relevance of dissemination architecture beyond individual metastatic sites or lesion morphologies.

The prognostic significance of MDP extends beyond surgical clearance. Complete macroscopic cytoreduction remains central to advanced ovarian cancer treatment,[5,34] yet MDP-high patients experienced substantially shorter survival after R0 resection (adjusted HR, 1.96). Occult microscopic disease may partly account for this survival difference, with emerging evidence suggesting that its fate depends on more than residual tumour burden. Recent studies in colorectal and liver cancers have revealed distinct malignant-cell states and immune niches associated with micrometastatic persistence and outgrowth.[35,36] Experimental evidence further demonstrates that systemic inflammation can reactivate dormant disseminated cells, while platelet-derived signals can suppress T-cell surveillance during metastasis.[37,38] In this context, MDP-associated inflammatory and platelet changes, together with a plasma cfRNA platelet-activation programme, complement previous studies of circulating markers in platinum resistance.[39,40] The persistent prognostic association of MDP may therefore be consistent with both occult tumour persistence and broader systemic tumour–host interactions, rather than reflecting surgical difficulty alone.

Our multi-omic findings suggest that MDP may involve complementary processes of peritoneal colonisation and immune escape. The stromal-remodelling axis and adhesion programmes in ascites spheroids are consistent with recent studies identifying mesothelial-containing spheroids and fibroblast–macrophage–tumour assemblies that facilitate peritoneal metastasis.[17,41,42] Notably, attenuation of interferon-response and HLA antigen-presentation programmes was shared across heterogeneous genomic backgrounds. Although HGSOC exhibits distinct HRD- and WGD-associated evolutionary trajectories,[19] WGD was enriched in MDP-high tumours across both HRD and HRP backgrounds. Among WGD-negative tumours, candidate HLA-I allele-copy loss was more frequent in MDP-high disease, consistent with genetic immune escape previously described in HGSOC.[15,20] In contrast, reduced HLA-I programmes persisted in WGD-positive tumours without detectable HLA-I loss, suggesting a contribution from copy-number-independent regulation. Previous studies have linked WGD to STING1 repression[20] and epigenetic silencing of antigen-presentation regulators,[43] while our methylation findings provide additional evidence of immune-related regulatory alterations. The experimental reversibility of PRC2-mediated MHC-I repression[44] raises the possibility that some components of MDP-associated immune attenuation may be therapeutically reversible, although this remains to be established.

Beyond malignant-cell immune attenuation, our analysis resolved six tumour-microenvironment ecosystems spanning distinct myeloid, T-cell and vascular–stromal states. Among these, TME-Angio showed higher activity in MDP-high tumours and combined ESM1+/APLN+ angiogenic endothelial cells and RGS5+ pericytes with MMP11+ myCAFs and C1QC+ macrophages. Angiogenic tip-cell programmes have been implicated in tumour progression,[45] while MMP11+ myCAFs resemble previously characterised POSTN-rich, TGF-β-responsive fibroblast states associated with ovarian cancer invasion and reduced CD8+ T-cell infiltration.[42,46-49] Recent studies further implicate fibroblast–myeloid interactions in peritoneal metastatic colonisation and T-cell dysfunction.[17] Notably, TME-Angio activity was inversely associated with malignant-cell MHC-I programmes, even after adjustment for MDP status. In external spatial sections, this ecosystem was associated with myeloid-enriched regions and lower T-cell abundance, while multiplex imaging of MDP-annotated tumours revealed reduced intratumoural CD8+ infiltration relative to the stroma. These findings are consistent with spatial studies showing that fibroblast neighbourhoods, malignant-cell antigen presentation and lymphocyte localisation jointly shape tumour immune organisation.[18,50-52] Our findings build on this understanding by linking malignant-cell immune attenuation and altered lymphocyte distribution to a defined macroscopic dissemination phenotype.

The clinical relevance of MDP lies in its potential for preoperative risk stratification and for informing future treatment strategies. The attenuated antigen-presentation programmes and reduced intratumoural T-cell infiltration associated with MDP are particularly relevant in light of recent advances in ovarian cancer immunotherapy.[53,54] Whereas adding atezolizumab to first-line chemotherapy and bevacizumab failed to improve progression-free survival in IMagyn050,[53] the phase III KEYNOTE-B96 trial demonstrated progression-free and overall-survival benefits from pembrolizumab plus weekly paclitaxel, with or without bevacizumab, in platinum-resistant recurrent disease.[54] Although neither trial evaluated MDP, these findings highlight the importance of understanding the tumour immune states associated with therapeutic benefit. Antigen-presentation escape in an ovarian cancer neoantigen-vaccine trial[55] and experimental restoration of immune programmes through PRC2 or DNMT–HDAC6 inhibition[43,44,56] further support investigating the reversibility of MDP-associated immune attenuation, without assuming that reduced HLA-I programmes necessarily confer checkpoint resistance. The feasibility of preoperative MDP prediction was further supported by primary-tumour MRI. Primary-tumour diffusion-weighted MRI predicted MDP across four external centres (pooled AUC, 0.814), and MRI-predicted MDP remained associated with survival after adjustment for established clinical variables. By predicting from primary-tumour MRI a dissemination phenotype associated with WGD enrichment and immune attenuation, our approach extends previous radiogenomic studies[24,57] and suggests a path towards biologically informed risk assessment in the preoperative setting. The clinical relevance of this approach may be particularly important in early-stage ovarian cancer or in patients with limited apparent disease dissemination, where aggressive metastatic behaviour cannot yet be inferred from conventional imaging. Primary-tumour MRI may capture intrinsic tumour characteristics associated with a propensity for miliary–diffuse dissemination before it becomes clinically apparent. Future studies should specifically investigate this hypothesis and determine whether such imaging signatures could enable earlier risk stratification and inform personalised treatment strategies.

This study has several limitations. Our dissemination atlas was reconstructed from retrospective clinical records with variable anatomical coverage at a single centre, and the reproducibility of the four patterns under different ascertainment conditions and across independently annotated cohorts remains to be established. Residual confounding by disease burden, treatment era and maintenance therapy remains possible. Survival was measured from surgery, a different treatment landmark for PDS and IDS. The molecular and spatial analyses involved substantially smaller patient subsets; candidate HLA-I allele-copy losses were inferred from tumour-only sequencing, and external spatial datasets lacked MDP annotations. The MRI cohorts were restricted to patients undergoing primary debulking surgery, with manual primary-tumour segmentation and relatively small samples at individual external centres. External calibration showed overestimation of MDP-high probability, and direct MRI prediction of molecular or immune features was not established. Longitudinal and functional studies are needed to determine whether malignant-cell immune attenuation and vascular–stromal organisation contribute to miliary dissemination or emerge during metastatic progression, and whether the regulatory suppression of antigen presentation can be reversed. Prospective multicentre studies are needed to establish the incremental clinical value of MRI-based MDP prediction over conventional imaging assessments and to investigate its relationship with molecular phenotypes. Ultimately, understanding these relationships may help explain why distinct patterns of metastatic spread carry different risks even after complete macroscopic tumour clearance.

## Main References

## Figure legends

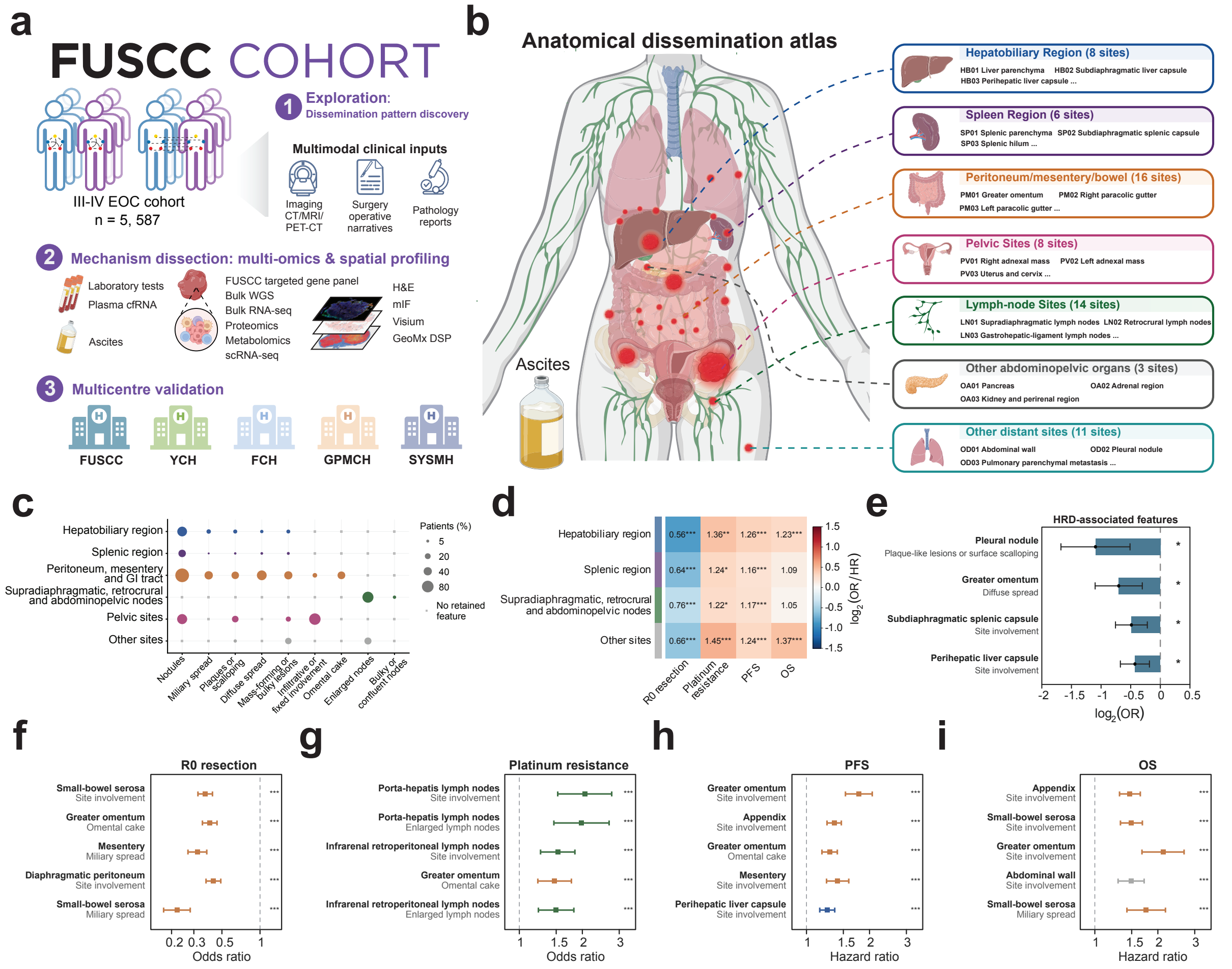

**Fig. 1 | A patient-level atlas of ovarian cancer dissemination.** **a**, Study overview. **b**, Anatomical framework comprising 66 sites in seven modules. **c**, Patient-level frequencies of lesion morphologies across anatomical modules (n = 5,587). **d**, Univariable associations of four anatomical groups with complete macroscopic resection (R0), primary platinum resistance, progression-free survival (PFS) and overall survival (OS). Colours indicate $\log_2$-transformed odds ratios (ORs) or hazard ratios (HRs). **e**, Associations of dissemination features with homologous-recombination deficiency (HRD) versus proficiency (HRP). **f–i**, Five associations with the smallest two-sided Wald P values for R0 (**f**), platinum resistance (**g**), PFS (**h**) and OS (**i**). Benjamini–Hochberg-adjusted P values were calculated; *q < 0.05, **q < 0.01, ***q < 0.001. Abbreviations: FUSCC, Fudan University Shanghai Cancer Center; EOC,

epithelial ovarian cancer; WGS, whole-genome sequencing; cfRNA, cell-free RNA; scRNA-seq, single-cell RNA sequencing; mIF, multiplex immunofluorescence; DSP, digital spatial profiling; GI, gastrointestinal; YCH, Yunnan Cancer Hospital; FCH, Fujian Cancer Hospital; GPMCH, Gansu Provincial Maternity and Child Care Hospital; SYSMH, Sun Yat-sen Memorial Hospital.

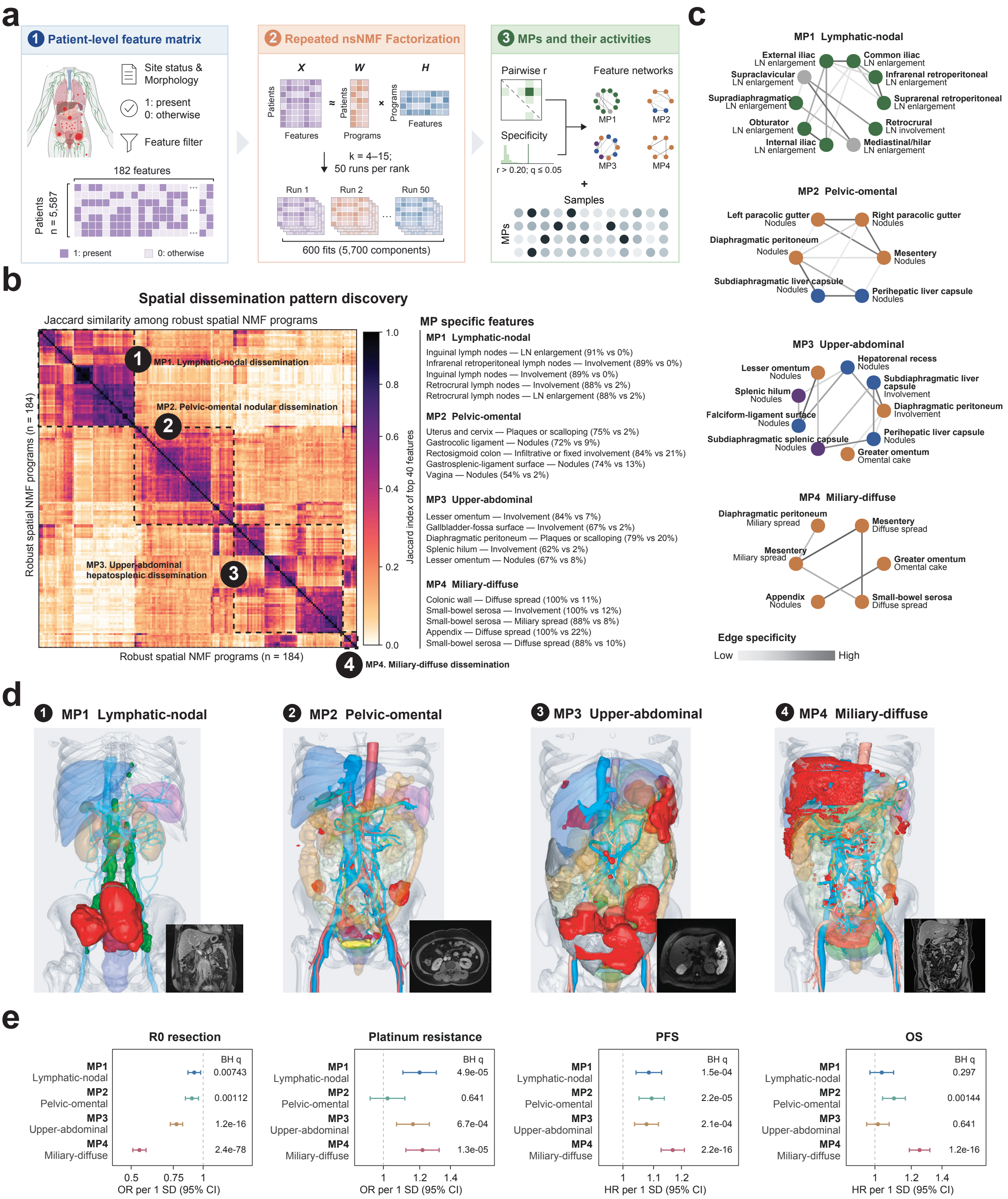

**Fig. 2 | Four coexisting patterns of ovarian cancer dissemination. a,** Discovery of dissemination patterns by non-negative matrix factorisation (NMF) and calculation of four continuous pattern scores per patient by non-negative least squares (NNLS), using 182 dissemination features from 5,587 patients. **b,** Heat map of Jaccard similarity among the top 40 features of 184 robust NMF programmes grouped into four consensus patterns. Percentages

give feature frequencies within each pattern versus all other programmes. **c,** Pattern-specific feature co-occurrence networks. Node colours indicate anatomical regions; edge shading denotes specificity. **d,** Representative three-dimensional (3D) anatomical reconstructions and corresponding MRI images for the four dissemination patterns. **e,** Univariable clinical associations per standard-deviation increase in each pattern score. Benjamini–Hochberg-adjusted P values were calculated across the 16 pattern–endpoint associations.

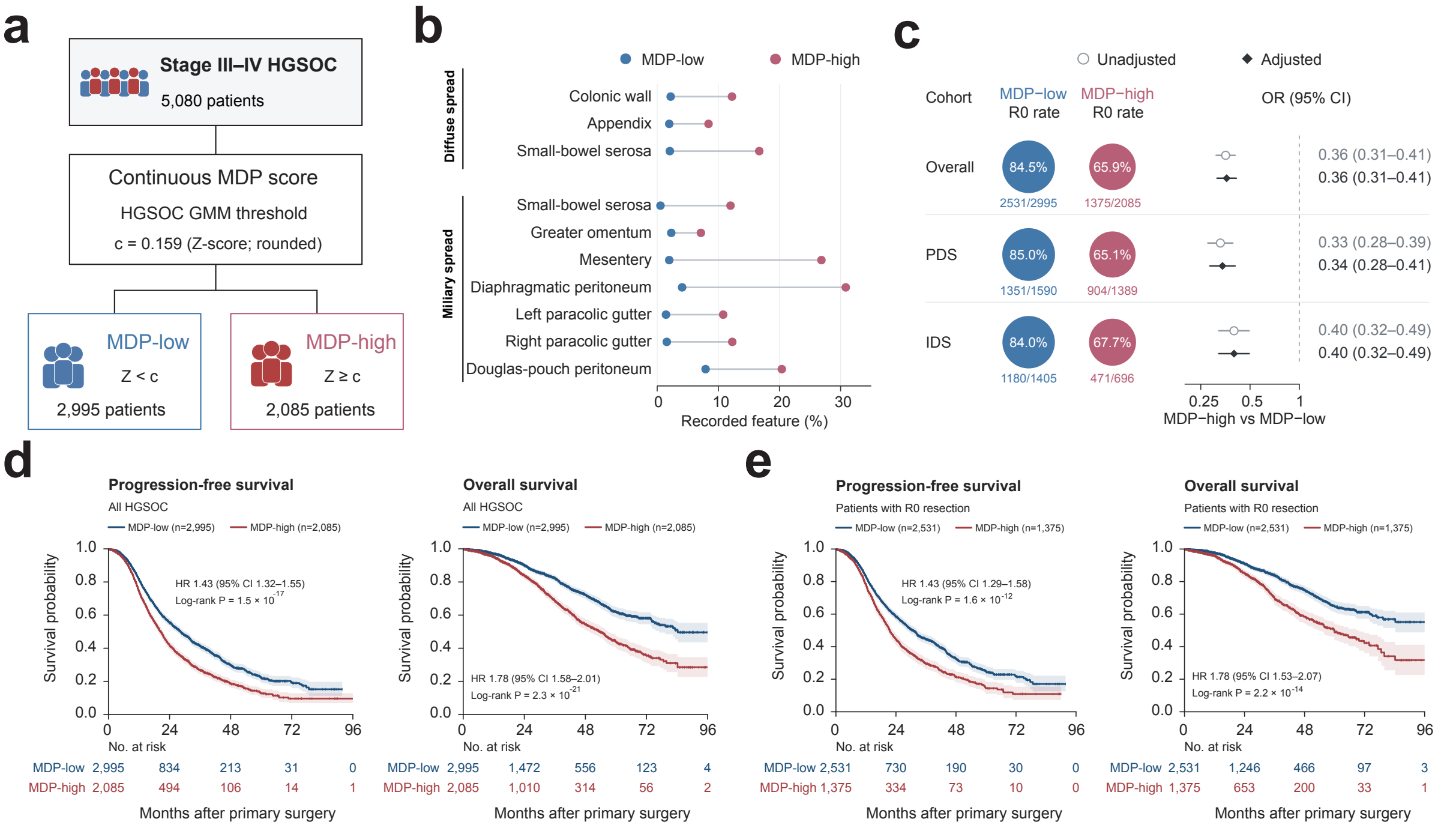

**Fig. 3 | MDP remains associated with adverse outcomes after complete resection. a,** Classification of 5,080 patients with high-grade serous ovarian cancer (HGSOC) into MDP-high and MDP-low groups using a Gaussian mixture model (GMM) fitted to continuous MDP scores in the HGSOC cohort. **b,** Frequencies of ten MDP-associated dissemination features in MDP-high and MDP-low patients. **c,** Complete macroscopic resection (R0) rates and unadjusted (open circles) and adjusted (filled diamonds) logistic odds ratios (ORs) for MDP-high versus MDP-low patients in the full HGSOC cohort (n = 5,080); models adjust for age and FIGO stage, and additionally for surgical pathway in the full cohort; bars indicate 95% Wald confidence intervals (CIs). **d,e,** Kaplan–Meier estimates of progression-free survival (PFS) and overall survival (OS) in the full HGSOC cohort (**d**) and R0 subgroup (**e**).

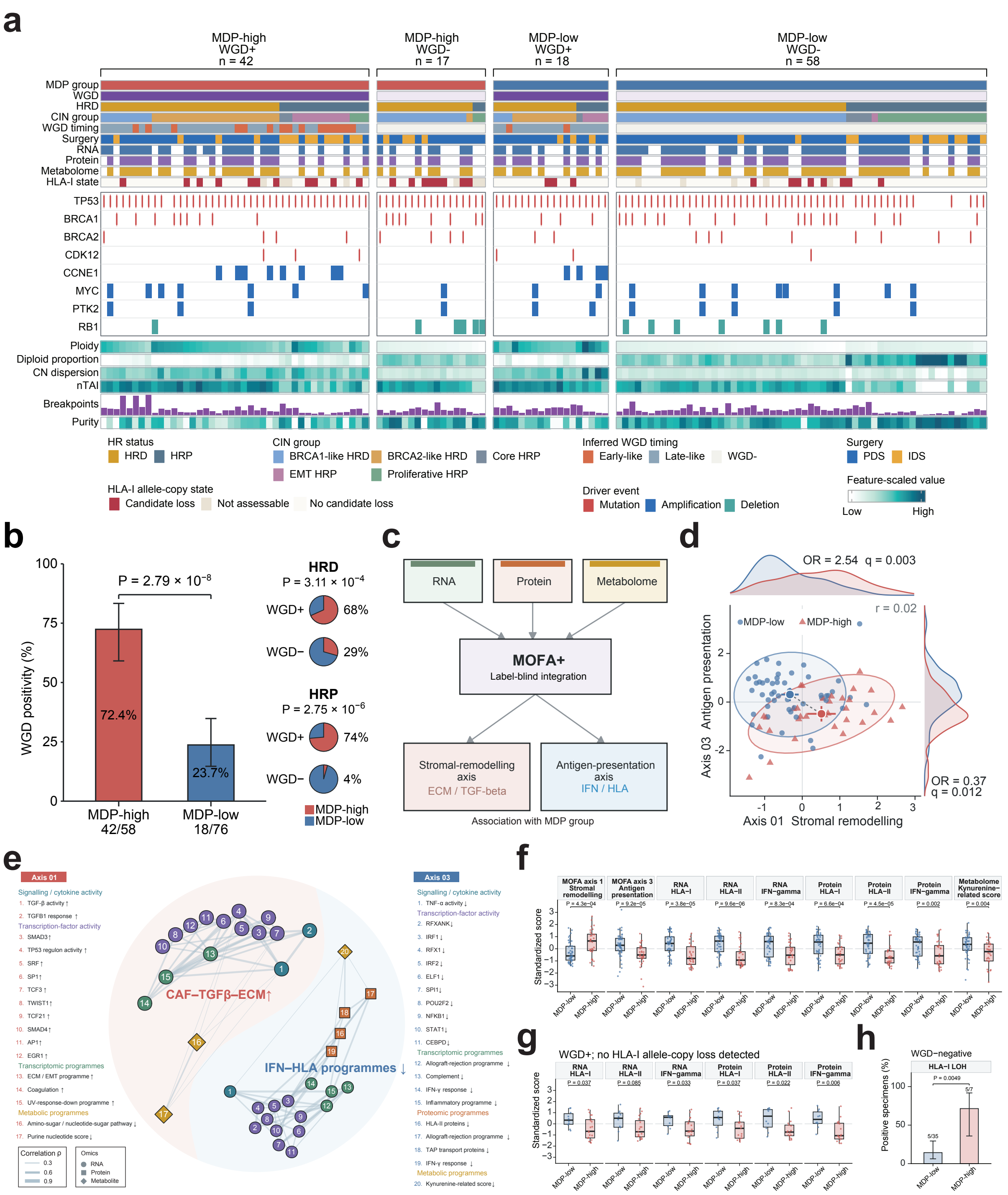

**Fig. 4 | Genomic and multi-omic features of MDP-associated immune attenuation. a,** Genomic and assay annotations for 135 specimens from 134 patients, grouped by miliary–diffuse dissemination pattern (MDP) and whole-genome doubling (WGD). Continuous variables are independently min–max scaled; gene rows show tumour-only variant and copy-number calls. **b,** Patient-level WGD prevalence by MDP status and MDP-high proportions

stratified by WGD and homologous-recombination status; exact binomial 95% confidence intervals (CIs) and two-sided Fisher exact tests. **c,** Multi-omics factor analysis (MOFA+) of 82 tumours, without MDP or WGD labels as inputs. **d,** Stromal-remodelling and antigen-presentation axes by MDP status. Curves show densities; ellipses enclose 80% data regions; large symbols and crosshairs indicate means and 95% t CIs. Odds ratios are per standard deviation; r, Pearson correlation. **e**, Axis-associated programmes: shapes indicate assays, colours programme families, edge widths positive Spearman correlations and arrows MDP-associated directions. **f,** MOFA+ axis scores, immune-programme scores and kynurenine-related scores in MDP-high and MDP-low tumours (RNA, n = 82; protein, n = 78). **g,** Immune-programme scores in WGD-positive specimens without detectable candidate HLA-I allele-copy loss at assessable loci (RNA, n = 29; protein, n = 27). **h,** Candidate HLA-I allele-copy loss in evaluable WGD-negative specimens (5/7 MDP-high versus 5/35 MDP-low); Wilson 95% CIs and two-sided Fisher exact test. P values were adjusted using the Benjamini–Hochberg procedure.

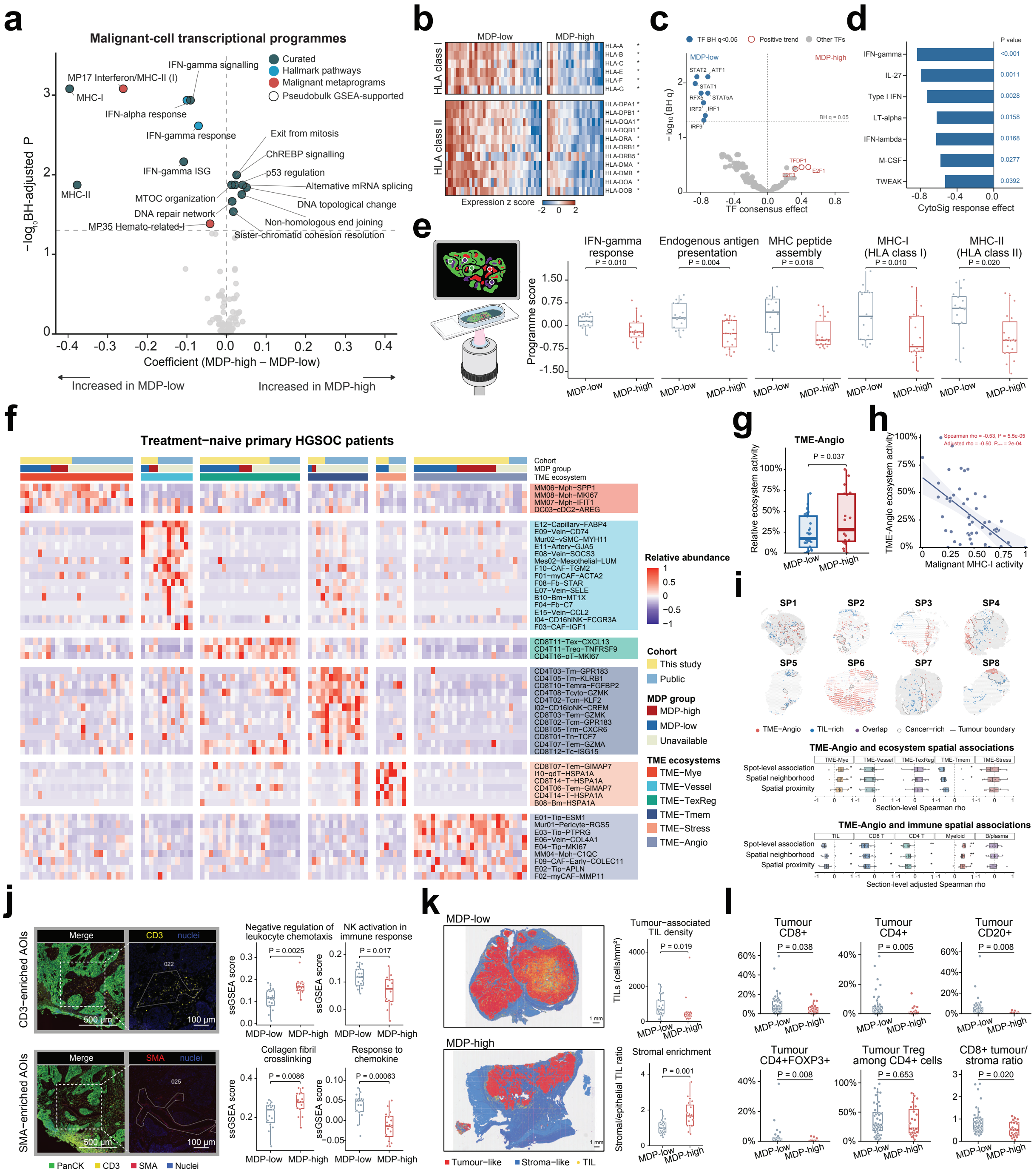

**Fig. 5 | Malignant-cell immune attenuation and vascular–stromal organisation in MDP. a–d,** Malignant-cell profiles from 57 patients. **a,** MDP-associated transcriptional programmes; outlines indicate concordant pseudobulk enrichment (q < 0.05). **b,** Heat map of HLA gene expression (*q < 0.05; voom–limma). **c,** Consensus transcription-factor effects. **d,** CytoSig-inferred cytokine-response effects associated with MDP status. **e,** GeoMx digital spatial profiling (DSP) of pan-cytokeratin (PanCK)-enriched areas of interest (AOIs); AOI-level

Wilcoxon rank-sum tests. **f,** Heat map of the relative abundance of 49 cell states across 108 treatment-naive primary HGSOC patients, grouped by six tumour-microenvironment ecosystems. Colours indicate scaled relative abundance. **g**, Box plots of relative TME-Angio ecosystem activity in MDP-high (n = 19) and MDP-low (n = 32) patients. The difference in group means was assessed using a two-sided patient-label permutation test (100,000 permutations). **h,** Association with malignant-cell MHC-I; line and shading indicate linear fit and 95% confidence interval; Spearman and MDP-adjusted partial Spearman correlations are shown, with permutation testing for the latter. **i,** Spatial distributions of TME-Angio and lymphocyte-rich domains in eight Visium sections. Section-level associations with other ecosystems and immune-cell populations are shown. Immune-cell associations adjust for total immune abundance; exact two-sided Wilcoxon signed-rank tests (*$P < 0.05$, **$P < 0.01$). **j,** Representative DSP images and transcriptional programme scores in CD3- and smooth muscle actin (SMA)-enriched AOIs (CD3, 12 MDP-high and 18 MDP-low; SMA, 16 MDP-high and 15 MDP-low). **k,** Histological tumour-infiltrating lymphocyte (TIL) densities and stromal-to-epithelial TIL ratios in 42 patients. **l**, Multiplex immunofluorescence (mIF) analysis of lymphocyte subsets and their tumour–stroma distribution in 63 patients.

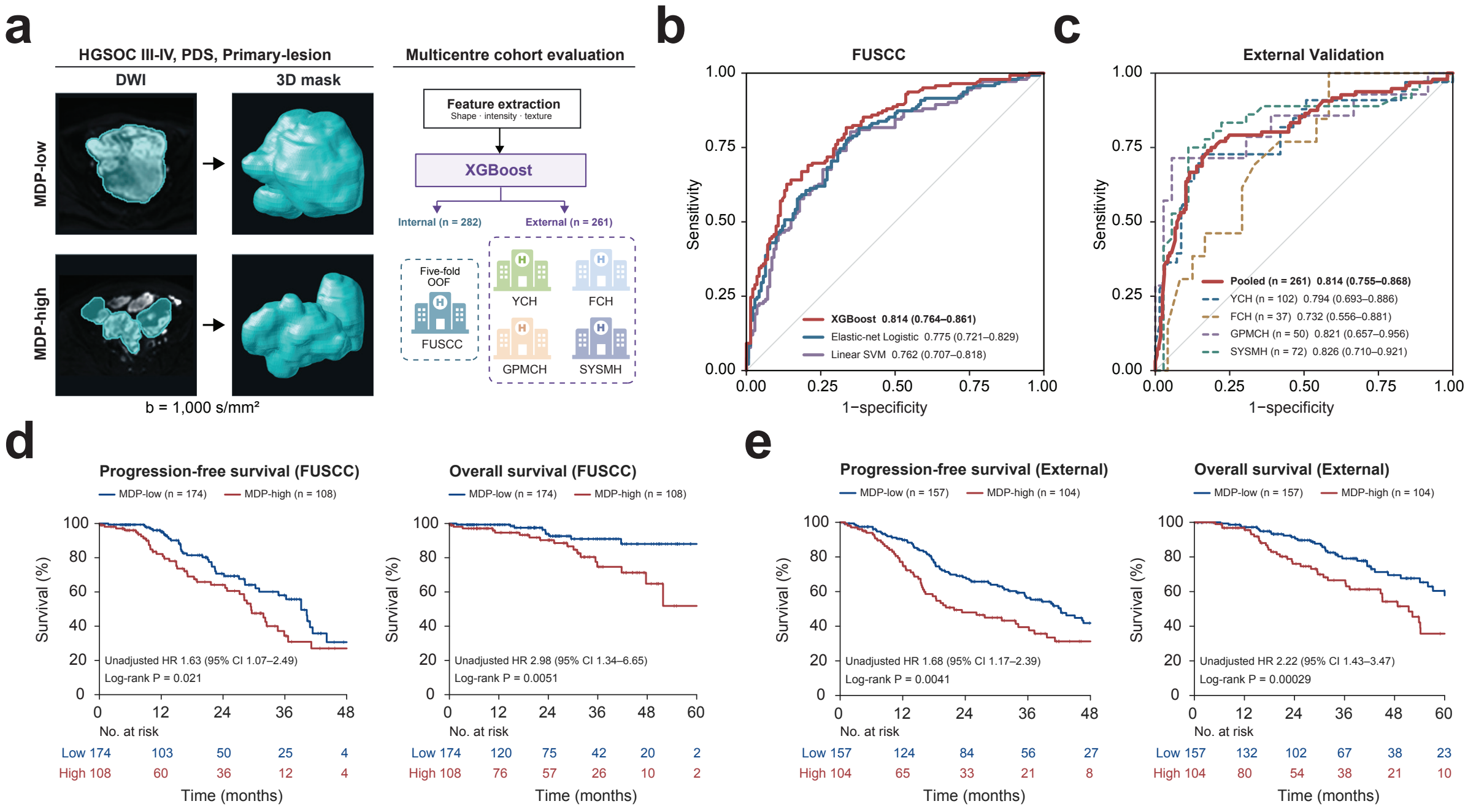

**Fig. 6 | Preoperative primary-tumour MRI predicts MDP and stratifies survival. a,** Radiomics workflow and examples of primary-tumour diffusion-weighted MRI (b = 1,000 s mm$^{-2}$). **b,** Receiver-operating-characteristic (ROC) curves from held-out predictions from five outer folds in 282 FUSCC patients. **c,** Performance of the frozen XGBoost model in 261 external patients from Yunnan Cancer Hospital (YCH; n = 102), Fujian Cancer Hospital (FCH; n = 37), Gansu Provincial Maternity and Child Care Hospital (GPMCH; n = 50) and Sun Yat-sen Memorial Hospital (SYSMH; n = 72). **d,e,** Kaplan–Meier progression-free survival (PFS) and overall survival (OS) by MRI-predicted miliary–diffuse dissemination pattern (MDP) in the internal cohort (**d**) and pooled external cohort (**e**).

## Methods

### Study cohorts

***Ethics.*** The study followed the Declaration of Helsinki. Ethics approval for the study, including multicentre retrospective imaging validation, was obtained from Fudan University Shanghai Cancer Center (FUSCC; 2601-Exp365, 2508-Exp538 and 2509329-7), Shanghai First Maternity and Infant Hospital (SFMIH; KS25394), and the Obstetrics and Gynecology Hospital of Fudan University (FOGH; 2025-152). Consent was waived for retrospective analyses of de-identified data; written informed consent was obtained for prospective sample collection.

***Clinical cohort.*** The discovery cohort comprised 5,587 patients with FIGO stage III–IV epithelial ovarian cancer who underwent surgery at FUSCC between January 2018 and March 2026, including 5,080 with HGSOC. Imaging, operative and pathological records were linked to clinical outcomes. Dissemination patterns were derived in the full cohort, with subsequent MDP classification and clinical analyses performed in all 5,080 patients with HGSOC (Supplementary Tables 1 and 11).

***Molecular cohorts.*** Institutional specimens came from FUSCC, SFMIH and FOGH. WGS covered 135 specimens from 134 participants. Within this series, 82 FUSCC PDS tumours contributed matched multi-omic profiles: RNA for all 82, and protein and metabolite measurements for 78 each. Matched single-cell profiles from 57 FUSCC patients supported malignant-cell analyses. Specimen codes, surgical pathways and assay participation are listed in **Supplementary Table 15**. Institutional and public datasets formed the single-cell reference atlas (**Supplementary Table 18**).

***MRI cohorts.*** The MRI series included 282 FUSCC patients and 261 patients from Yunnan Cancer Hospital, Fujian Cancer Hospital, Gansu Provincial Maternity and Child Care Hospital and Sun Yat-sen Memorial Hospital. All had FIGO stage III–IV HGSOC treated with PDS (**Supplementary Fig. 5** and **Supplementary Table 27**). The median interval between MRI and surgery was 4 days (IQR, 2–7 days). Patients with inadequate clinical records or MRI image quality were excluded. Preoperative pelvic MRI was performed using 1.5-T or 3.0-T scanners from Siemens, GE Healthcare and Philips. The imaging protocol included T2-weighted imaging, diffusion-weighted imaging (DWI), apparent diffusion coefficient (ADC) maps and contrast-enhanced T1-weighted imaging. High b values were 800 or 1,000 s $mm^{-2}$. Centre-specific DWI acquisition parameters are summarised in **Supplementary Table 28**. Reference MDP labels were derived from integrated imaging, operative and pathological records. The four external cohorts were assigned MDP labels using the same record-review procedure, 182-feature encoding, fixed NMF consensus patterns, NNLS projection and GMM-derived threshold as the discovery cohort, independently of MRI model predictions.

## Dissemination pattern discovery

***Record extraction and review.*** Patient-grouped Chinese imaging, operative and pathological records were processed with institutionally deployed gpt-oss-120b. For patients undergoing IDS, dissemination profiles were based on findings documented before the initiation of neoadjuvant chemotherapy. Zero-shot prompts requested anatomical involvement, morphology, evidence states and source quotations in structured JSON. The dictionary comprised 66 anatomical sites and nine non-exclusive descriptors (**Supplementary Tables 3 and 4**). Two radiologists independently reviewed AI extractions for all 5,587 patients, with disagreements adjudicated by a third reviewer. All analyses used the clinician-corrected annotations (**Supplementary Tables 2 and 5–7**).

***Feature encoding.*** Binary features represented site involvement and site–morphology combinations, with morphology retained only at involved sites. Pattern discovery excluded primary adnexal sites, features with prevalence below 1% and site-status features with prevalence above 60%, retaining 54 site-status and 128 morphology features. The extraction schema distinguished present, absent, uncertain and not assessed. In the primary analysis, recorded positive findings were encoded as 1, whereas absent, uncertain and not-assessed findings were encoded as 0.

***Robust NMF.*** Non-smooth NMF[58] was applied to the patient-by-feature matrix at ranks 4–15, with 50 initialisations per rank, smoothing parameter 0.5 and at most 160 multiplicative-update iterations, yielding 600 fits. Each component was represented by its 40 highest-weighted features. Fits with the same replicate index across ranks formed an initialisation stream. Components were screened for reproducibility across ranks and streams and for non-redundancy within each stream.[59] Cross-rank reproducibility required at least 28 shared features (70%) with a component at another rank in the same stream; cross-stream reproducibility required at least eight shared features (20%) with a component from another stream. Eligible components were ranked first by cross-stream support and then by cross-rank support; maximal overlaps and lower reconstruction error resolved remaining ties. Components were selected in descending order, removing candidates that shared more than eight features with an already selected component in that stream. This procedure retained 184 robust components.

***Consensus patterns.*** Robust components were grouped by iterative greedy clustering of their feature sets.[59] For each unassigned component, we counted other components sharing at least 20 of its 40 features. The component with the most such matches founded a cluster, provided that at least four other components met this criterion. The unassigned component with the greatest overlap with the current consensus was then added if it shared at least 13 features. After each addition, the consensus was updated to the 40 features occurring in the most cluster members, with ties resolved by the maximum NMF loading across those members. Cluster growth stopped when no remaining component met the inclusion threshold. Assigned components were removed, and founder selection was repeated until no eligible founder remained. The four groups contained 56 (MP1), 57 (MP2), 63 (MP3) and 8 (MDP) components. Clinical outcomes were not used in discovery.

***Patient scores and stability.*** Group basis vectors were the medians of member components' L2-normalised loadings across 182 features, normalised to unit L1 sum. Patient profiles were jointly projected onto the four vectors by non-negative least squares. Scores were standardised to the 5,587-patient discovery cohort. Gaussian mixture models with one to six components were fitted to standardised MDP scores in the cohort of 5,080 patients with HGSOC. Among multicomponent fits, the two-component model maximised out-of-bag classification stability. Equal posterior probabilities defined the MDP-high threshold (raw score ≥6.243582; standardised score ≥0.159349). Stability was assessed by resampling components within the four groups 1,000 times, rebuilding the basis and comparing scores and feature profiles. Additional NMF refits and score-sensitivity analyses are described in **Supplementary Methods** and **Supplementary Figs. 2–4**.

***Feature co-occurrence networks.*** Within each pattern, features were ranked by mean L1-normalised component loading multiplied by (0.5 + 0.5s), where s was the fraction of components placing that feature in the upper 5% of loadings. The top ten features formed candidate nodes after excluding site-status features when a morphology feature from the same site was selected. Pearson correlations were calculated for all 16,471 feature pairs, with BH correction. Pair specificity was the fraction of correlations involving either feature that did not exceed the target correlation.[60] Edges required $r > 0.20$, $q \leq 0.05$ and specificity ≥344/361. Connectivity, defined as observed divided by possible edges, was compared with 10,000 size-matched random feature sets using one-sided permutation tests and BH correction across four patterns.

## Clinical and plasma analyses

***Endpoints.*** R0 denoted no macroscopic residual disease. PFS was measured from cytoreductive surgery to disease progression or death from any cause, and OS from cytoreductive surgery to death from any cause. Patients without an event were censored at their last available follow-up. Primary platinum resistance denoted progression during platinum treatment or within six calendar months after completion of first-line platinum therapy.

***Association models.*** Logistic regression assessed R0 and platinum resistance; Cox models assessed survival, using Efron's method for ties. Effect estimates used 95% Wald confidence intervals. Regional, individual-feature and separate pattern-score analyses were univariable. The clinical feature screen included the 182 NMF features plus greater-omental and mesenteric involvement. Benjamini–Hochberg correction covered 16 regional comparisons (four categories × four endpoints), 736 feature comparisons (184 × four) and 16 pattern-score comparisons. For each endpoint, the five features with the smallest two-sided Wald P values were displayed. Continuous scores were tested per discovery-cohort standard deviation. Joint models included all four scores, age per ten years, FIGO stage, surgical pathway and HGSOC histology; platinum-resistance and survival models additionally included R0. Adjusted regional models included age, FIGO stage, surgical pathway and histology, with additional adjustment for R0 in platinum-resistance and survival analyses (**Supplementary Table 8**).

***MDP groups and blood measurements.*** HGSOC analyses used continuous MDP scores or binary MDP classifications. Models adjusted for age, FIGO stage and surgical pathway, with additional adjustment for R0 in full-cohort platinum-resistance and survival analyses. Survival analyses were also performed separately by surgical pathway and among patients achieving R0. Additional models included recorded miliary spread, diffuse spread and mesenteric involvement, including the mesenteric root (**Supplementary Tables 12–14**). Proportional hazards were assessed using Schoenfeld residuals. Preoperative blood measurements were compared in the PDS subset using Wilcoxon rank-sum tests, corrected across 12 analytes.

***Plasma cell-free RNA.*** Plasma cfRNA profiles came from our previously published FUSCC cohort.[40] This analysis included 157 patients (51 MDP-high and 106 MDP-low). Genes were filtered with edgeR filterByExpr (minimum count, 10; minimum total count, 15) and TMM-normalised.[61] Voom–limma models included MDP group, age, FIGO stage and sequencing batch, with repeated profiles grouped by patient.[62,63] Moderated t statistics for 6,343 genes supplied fgseaMultilevel rankings (gene-set size, 15–500; eps = 0; sampleSize = 101; nPermSimple = 10,000; seed = 20261022).[64] Gene-set enrichment was assessed across 3,057 gene sets, with Benjamini–Hochberg correction across the full testing set.[59,65]

***Ascites.*** Ascites presence, volume and cytology were examined in patients undergoing PDS across MDP-score quartiles defined in the discovery cohort. Associations with continuous MDP scores were assessed using Wilcoxon rank-sum tests for ascites presence and cytology and Spearman correlation for ascites volume, with Benjamini–Hochberg correction across three comparisons (**Extended Data Fig. 6**).

**Sample collection and profiling**

***Sample collection.*** Fresh primary-tumour tissue was collected for single-cell RNA sequencing and paired molecular profiling. Bulk-assay specimens were generally frozen and stored in liquid nitrogen. For OCT-embedded specimens, tissue free of embedding medium was excised before nucleic-acid extraction.

***Bulk RNA sequencing.*** TRIzol-extracted RNA underwent quantity and integrity assessment and oligo(dT) enrichment. BGI Optimal Dual-mode mRNA libraries were circularised, converted to DNA nanoballs and sequenced as paired-end reads on DNBSEQ. SOAPnuke v2.3.2[66] removed adaptor contamination, reads with more than 5% ambiguous bases and reads with more than 40% of bases below Phred 20. Bowtie2 v2.4.5 aligned reads to the reference transcriptome; RSEM v1.3.1 quantified transcripts.[67,68]

***Proteomics.*** Tissue was lysed in 5% sodium deoxycholate and 100 mM Tris-HCl, pH 8.5, reduced with dithiothreitol and alkylated with iodoacetamide. Trypsin digestion was performed using an enzyme-to-protein ratio of 1:50 for 15–18 h at 37 °C. Desalted peptides underwent data-independent acquisition on an Orbitrap Astral coupled to Vanquish Neo liquid chromatography. MS1 scans covered m/z 380–980 at resolution 240,000 at m/z 200; MS2 used 299 windows of 2 m/z. DIA-NN v1.8.1 allowed one missed cleavage, fixed cysteine carbamidomethylation, and variable methionine oxidation and protein N-terminal acetylation.[69] Protein identification was controlled at 1% FDR.

***Metabolomics.*** Tissue homogenates were extracted with methanol–acetonitrile (1:1, v/v). Supernatants were dried and reconstituted in acetonitrile–water (1:1, v/v). Separation used a BEH Amide column (2.1 × 100 mm, 1.7 µm), with positive- and negative-ion acquisition on an Orbitrap Exploris 480. ProteoWizard converted raw files and XCMS generated peak-area matrices.[70,71] Putative annotations combined accurate mass (25-ppm tolerance) and MS/MS matching to an in-house library. Proteomics and metabolomics were performed by Shanghai Applied Protein Technology Co., Ltd.; each assay included pooled quality-control injections after every 12 study samples.

***Single-cell RNA sequencing.*** Fresh tissue was minced and dissociated enzymatically and mechanically at 37 °C, filtered through 70-µm and 30-µm strainers, and subjected to erythrocyte lysis. Acridine orange/propidium iodide staining assessed viability; suspensions below 80% underwent Miltenyi Biotec dead-cell removal. Chromium X and Chromium Next GEM Single Cell 3′ Reagent Kit v3 were used for library preparation, followed by paired-end sequencing and Cell Ranger quantification. Public datasets were obtained as deposited count matrices.

**Genomic and multi-omic analyses**

***WGD and HRD.*** Tumour WGS was processed in tumour-only mode with the nf-core oncoanalyser pipeline v2.3.0 and the HMF GRCh38 reference. PURPLE v4.3 integrated AMBER v4.2 B-allele frequencies, COBALT v2.2 read-depth ratios and structural-variant breakpoints to estimate purity, ploidy and allele-specific copy number. WGD status used the PURPLE wholeGenomeDuplication output.[72] ovaHRDscar v0.1.0 was applied to PURPLE segments of at least 100 kb; scores ≥54 defined HRD and lower scores defined HRP.[73] WGD prevalence comparisons were performed at the patient level (n = 134; Supplementary Table 16).

***HLA allele-copy loss.*** LILAC provided HLA-A, HLA-B and HLA-C alleles and copy estimates.[74] Where reference harmonisation was required, reads from primary HLA intervals and chromosome 6 alternative contigs were realigned to HMF with BWA-MEM2. Candidate loss required minor-allele copy number <0.3 and major-allele copy number >0.7 at the same assessable, typed heterozygous HLA-I locus. Specimens homozygous at all typed loci could not be evaluated. Analyses without detected loss used available assessable loci; candidate-loss frequencies were compared by Fisher's exact test.

***Multi-omic integration.*** MOFA+ integrated RNA, protein and metabolite measurements from 82 FUSCC PDS tumours.[75] RNA underwent blind variance-stabilising transformation with DESeq2;[76] proteins were log2-transformed and sample-median-centred. Log2-transformed metabolite measurements were adjusted by feature-wise linear regression on run group. Proteins required observations in at least 75% of available specimens. Variance ranking retained 1,000 RNA, 1,000 protein and 600 metabolite features, each standardised across specimens, with missing values left unaltered. MDP and WGD labels were excluded from fitting.

***Factor selection and associations.*** Models used 8, 10, 12 or 15 factors and five seeds each; the primary model had ten factors and seed 42. Factors were matched by complete-linkage clustering of 1 − |Pearson r| at |r| ≥0.80. Consensus axes required representation in all five ten-factor runs and at least three factor-number settings, median |r| ≥0.90 to the representative, and median summed view-specific explained variance ≥2%. Seven axes qualified. Separate logistic models assessed the association of each standardised axis with MDP, with correction across seven axes. RNA, protein and metabolite programme associations informed interpretation. The two MDP-associated axes were subsequently examined in a joint logistic model adjusted for tumour purity, HRD and WGD status (**Supplementary Table 17**).

***Bulk immune programmes.*** HLA-I and HLA-II scores used intact MSigDB 2026.1.Hs sets GO:0042612 and GO:0042613. Gaussian-kernel GSVA was applied to full log2(TPM + 1) RNA and log-transformed protein matrices.[77] Wilcoxon rank-sum comparisons were corrected across nine displayed endpoints. WGD-positive comparisons were repeated in participants without detected HLA-I allele-copy loss at assessable loci.

## External WGD analyses

***Clonal timing and micronuclei.*** SPECTRUM single-cell genomes and phylogenies were used to assign broad 6p21/MHC-region allelic loss to branches relative to stable WGD events.[20] Same-branch events formed a separate category. Micronucleus measurements were compared at the patient level by Wilcoxon rank-sum testing.

***Experimental transcriptomes and chromatin accessibility.*** Published mouse mammary-tumour transcriptomes and scATAC-seq profiles were analysed by WGD status.[43] Time-course analyses used 12 tumours, three per WGD-by-time group at days 8 and 14. Five module-score interactions were tested by exact blocked permutation, corrected across modules. scATAC clusters with coordinated Epcam and keratin accessibility and low immune-lineage signals were selected as epithelial candidates. Gene-body and 2-kb upstream activity counts were summed by tumour in two WGD-positive and two WGD-negative tumours, filtered with filterByExpr and TMM-normalised. WGD comparisons used robust edgeR quasi-likelihood testing, corrected across 19,114 genes.

***Ovarian-cancer DMR enrichment.*** GSE292140 ovarian-cancer methylation data included 154 tumours (95 WGD-positive and 59 WGD-negative). Author-provided, platform-corrected β values were transformed to M values. Autosomal probes were retained if they passed the Zhou et al. universal mask, contained no within-probe SNP and had measurements in at least 95% of samples, yielding 290,201 probes.[78] Differentially methylated CpGs were tested using limma (lmFit and eBayes), modelling M values as a function of WGD status and tumour purity, both obtained from the source study.[63] Differentially methylated regions (DMRs) were identified using DMRcate with the same design (cpg.annotate, fdr = 0.05; dmrcate, lambda = 1,000, C = 2; hg19).[79] Hyper- and hypomethylated DMRs were tested separately for enrichment in the 50 MSigDB Hallmark gene sets using missMethyl gsaregion, with all tested probes as background and correction for the number of probes per gene.[80] P values were adjusted using the Benjamini–Hochberg procedure separately within each direction (**Extended Data Fig. 4f**).

***Immune-gene promoter methylation.*** STING1 and GBP4 were examined in TCGA-OV (27K, n = 386) and GSE292140 (450K, n = 154). Each gene used the quality-controlled probe nearest a transcription start site within −1,500 to +500 bp. TCGA-OV limma models on M values included WGD and ABSOLUTE purity; the same probes were evaluated using GSE292140 WGD contrasts. One-sided P values tested higher methylation in WGD-positive tumours. Immune-gene screening required TCGA-OV $P < 0.05$ and either Δmedian β ≥0.05 or an increase ≥0.10 in the fraction with $\beta > 0.3$. STING1 and GBP4 were selected for exploratory analysis after review of their immune functions; plots show raw $\beta$ values.

**Single-cell transcriptomic analyses**

***Quality control and preprocessing.*** Count matrices were harmonised to 21,812 genes shared by the GRCh38 Ensembl 84, Ensembl 93 and GENCODE v32/Ensembl 98 annotations.[60] Scanpy retained cells with 500–8,000 detected genes, 1,000–100,000 counts and <20% mitochondrial counts.[81] Scrublet identified doublets within each library at an expected rate of 0.05; libraries retaining fewer than 50 cells were excluded.[82] Expression was normalised to 10,000 counts per cell and log1p-transformed. For whole-atlas integration, 3,000 highly variable genes were selected with the cell_ranger method using dataset as the batch variable. Mitochondrial, ribosomal, immunoglobulin, T-cell-receptor and heat-shock genes were excluded from selection; scVI input retained raw counts.

***Integration and benchmarking.*** scVI v1.3.3 used dataset as the batch covariate, negative-binomial likelihood, 30 latent dimensions, two 128-unit layers and dropout 0.1.[83] Training used a 90:10 training–validation split, batch size 1,024, learning rate 0.001 and 20-epoch KL warm-up. Models ran for up to 500 epochs, with early stopping after ten epochs without improvement in validation ELBO. Seven configurations were benchmarked: PCA, BBKNN, Harmony, Scanorama, SCALEX, and scVI with dataset alone or dataset plus sample covariates. scIB metrics were evaluated in the entire atlas and 50,000-cell subsets, scaled across methods, and combined as 40% batch correction and 60% biological conservation.[84]

***Cell-state annotation.*** We performed at least two rounds of unsupervised cell clustering and annotation. Canonical markers and reference predictions established broad cell identities, followed by Leiden reclustering within lineages and marker-based subtype assignment. Malignant epithelial cells were identified using epithelial-lineage markers and large-scale chromosomal copy-number alterations inferred from single-cell transcriptomes using inferCNV. The first level of clustering was performed using the function scanpy.tl.leiden with resolution = 0.1.

We then identified differentially expressed genes (DEGs; $\log_2$-transformed fold changes > 1, FDR < 0.05, Student's t-test). Cell annotation was further supported by TOSICA and CellTypist.[85,86] CellHint related study annotations to published cell-state labels;[87] three representative markers per resolved state were summarised from the full gene matrix. The final atlas contained 2,871,618 cells from 682 samples across 30 datasets, including malignant cells and 104 non-malignant categories (**Supplementary Tables 18–20**). Subtype-mean log-normalised expression was hierarchically clustered using 1 − Pearson correlation and Ward.D linkage, excluding the unresolved category.

***Malignant-cell programmes.*** Comparisons used patient-level profiles for 57 FUSCC patients, 21 MDP-high and 36 MDP-low. The 112 signatures comprised Hallmark, malignant meta-programs, published ovarian-cancer and curated functional programmes (**Supplementary Table 21**). Scanpy score_genes used 25 expression bins, 50 control genes per bin and seed 0 on log-normalised expression; scores were averaged within patients. Group effects used ordinary least squares with HC3 robust standard errors, corrected across 112 programmes. Patient-summed malignant counts underwent edgeR TMM and voom–limma analysis; moderated t statistics supplied enrichment rankings for the same programmes. HLA comparisons were corrected across 17 genes.

***Regulatory and cytokine activity.*** CollecTRI–ULM, DoRothEA–VIPER and pySCENIC–AUCell inferred transcription-factor activity.[88-90] For 154 factors shared across methods, activities were standardised across patients within each method. Consensus effects averaged the three MDP-high minus MDP-low effects with equal weights. Inference used 10,000 synchronised patient-label permutations and correction across 154 factors (**Supplementary Table 22**). CytoSig scores used patient-level HC3-robust linear models with two-sided Wald tests, corrected across 43 readouts (**Supplementary Table 23**).[91]

***External tumour-cell RNA sequencing.*** Tumour-cell-enriched GSE293643 profiles underwent edgeR TMM and limma–voom analysis.[10] Comparisons of the original miliary and non-miliary spread groups adjusted for specimen type and blocked repeated specimens by patient; pooled ascites-cell/spheroid comparisons additionally distinguished these fractions. Moderated t statistics supplied fgseaMultilevel enrichment, corrected across ten focused pathways. A separate comparison of ascites single tumour cells and natural spheroids used nine paired patients with patient fixed effects, followed by enrichment across Hallmark, Reactome and GO biological-process collections.

**Microenvironment and spatial mapping**

***Cell-state composition.*** Treatment-naive, unsorted primary HGSOC profiles were aggregated by patient across samples and libraries. A minimum of 500 non-malignant cells per patient was required, yielding a 104-state matrix for 108 patients (61 institutional, 47 public). Fractions used all retained non-malignant cells as denominator.

***Ecosystem discovery and associations.*** Min–max-scaled fractions underwent Lee NMF at ranks 2–10 with 100 runs per rank.[92] Cophenetic correlation and consensus structure supported rank six, which was fitted with 200 initialisations. Robustness used 200 resamples of 86 patients without replacement, recalculating scaling and fitting ranks 2–10 with 200 initialisations per rank. Each NMF coefficient was divided by the sum across six ecosystems to define relative activity; the largest identified the dominant ecosystem (**Supplementary Table 25**). Mean relative ecosystem activities were compared between MDP groups in 51 labelled patients using two-sided patient-label permutation tests (100,000 permutations), with Benjamini–Hochberg correction across the six ecosystems. TME-Angio–MHC-I partial Spearman correlations adjusted for MDP; P values used 99,999 within-group permutations. MHC-I scores were rescaled to their observed 0–1 range for display. Heatmap colours used row-wise z scores of cell-state fractions divided by four and clipped to [−1, 1].

***Spatial projection.*** Cell2location[93] projected the reference into eight GSE211956 Visium sections[94] from primary ovarian tumours collected at interval debulking surgery following platinum- and taxane-based neoadjuvant chemotherapy, without MDP annotations. Ecosystem maps used NMF-weighted sums of posterior fifth-percentile cell-state abundances, divided by total non-malignant abundance. TME-Angio used its seven-state core of tip endothelial, pericyte, venous endothelial and fibroblast states. Core scores were scaled by pooled 95th percentiles; dominant domains required the highest score and a relative margin ≥0.20 over the second-highest. Lymphoid-rich regions were the within-section top 5% of relative CD4, CD8, innate T/NK and B/plasma abundance. Quantitative association analyses included all 104 states and the full ecosystem loadings. Within each section, Spearman correlations compared scores at the same spot, across six-neighbour averages and across exponentially distance-weighted fields (bandwidth, three adjacent-spot spacings). Immune-target abundances were rank-adjusted for total immune abundance. The eight section-level correlations were compared with zero using exact two-sided Wilcoxon signed-rank tests.

**Digital spatial profiling and histology**

***GeoMx profiling.*** Guangzhou Genedenovo Biotechnology Co., Ltd. profiled FFPE sections by GeoMx.[95] After antigen retrieval, sections received fluorescent morphology markers and RNA probes carrying photocleavable oligonucleotides. PanCK-, CD3- and SMA-enriched areas of interest (AOIs) were selected from fluorescence images. Ultraviolet-released oligonucleotides were collected by microcapillary aspiration and sequenced. GeoMx NGS Pipeline generated digital counts; quality control included yield, alignment, saturation and negative-probe signal. Grubbs-test probe outliers present in at least 20% of segments were removed globally, followed by Q3 normalisation.

***Spatial programme comparisons.*** PanCK analyses included 34 AOIs from ten patients. Five programme scores averaged gene-wise standardised log2(Q3 + 1) expression. AOI comparisons used Wilcoxon rank-sum tests. Patient medians were compared by the difference in group means across all 252 allocations of five versus five patients. CD3 and SMA analyses used ssGSEA on log2(Q3 + 1) with intact MSigDB Hallmark, Reactome and GO biological-process sets (alpha = 0.25, normalisation enabled, size 10–500). Correction covered 6,656 sets per compartment; selected comparisons report nominal P values (**Supplementary Table 24**).

***H&E analysis.*** A pretrained ovarian-cancer segmentation network combining multiple image magnifications generated tumour and non-tumour tissue masks from spatially sampled H&E tiles of 42 patients.[96] Non-tumour tissue defined the stromal compartment. HoVer-Net with MoNuSAC-trained weights identified lymphocyte nuclei; their centroids were registered to the tissue masks.[97,98] Tumour-associated lymphocyte density counted cells within the tumour mask and its 50-µm external margin, divided by tumour area. The stromal-to-epithelial density ratio compared the 50–1,000-µm stromal neighbourhood with the tumour-plus-margin region, using respective regional areas. Patient-level comparisons used Wilcoxon rank-sum tests.

***Multiplex immunofluorescence staining.*** Tissue sections underwent multiplex immunofluorescence staining with antibodies against CD8 (clone SP239, ab178089), CD68 (clone EPR20545, ab213363), CD20 (clone EP459Y, ab78237), FOXP3 (clone 236A/E7, ab20034), CD4 (clone EPR6855, ab133616) and pan-cytokeratin (clone C-11, ab7753), all from Abcam. Fluorescence detection used the Multiplex IHC Detection kit (TSA amplification; ab312827, Abcam). Nuclei were counterstained with DAPI.

***Multiplex image analysis.*** Images were analysed in HALO v3.6.4134 (Indica Labs) after excluding folds, staining artefacts, non-specific staining and blood-cell regions. HALO AI separated tumour and stroma; Highplex FL v4.2.14 segmented and phenotyped cells. Nuclear segmentation used 20 annotated 1 × 1-mm regions from five images, accepting nuclear areas of 11.3–571.7 $\mu m^2$ and roundness ≥0.3. Cytoplasmic regions extended 3 μm beyond nuclei. Cytoplasmic positivity thresholds were CD8 22.7679, CD68 21.5848, CD20 34.5143, CD4 26.183 and pan-cytokeratin 13.6607; nuclear FOXP3 used 23.616. Cell fractions used all cells in the respective compartment as denominator. Treg abundance among CD4+ cells used the CD4+FOXP3+/CD4+ ratio; the CD8 tumour-to-stroma ratio compared compartment-specific fractions. Six endpoints were compared between MDP groups in 63 patients by Wilcoxon rank-sum tests (**Supplementary Table 26**).

**MRI radiomics**

***Feature extraction.*** Primary tumours were manually delineated slice by slice on preoperative DWI by a radiologist with 14 years of experience in gynaecological imaging, blinded to MDP labels, using MITK software (version 2021.02), excluding cystic regions with guidance from multisequence MRI. For patients with bilateral ovarian masses, the larger lesion was selected for analysis. To assess intra- and inter-reader reproducibility, 30 randomly selected tumours were resegmented by the original radiologist and a second radiologist with 7 years of experience in gynaecological imaging. Images were resampled to 1.6796875 × 1.6796875 × 5 mm using B-spline interpolation; binary primary-tumour masks used nearest-neighbour interpolation. PyRadiomics v3.0.1 extracted 851 shape, first-order and texture features from original and wavelet-transformed images.[99] All centres used intensity normalisation, a three-standard-deviation outlier setting and bin width 25. Of the 851 extracted radiomic features, 795 with intraclass correlation coefficients (ICCs) greater than 0.80 were retained for subsequent analysis.

***Model development.*** Model development used stratified nested cross-validation in the 282-patient FUSCC cohort, with five outer folds and four inner folds. Within training folds, features with >20% missingness were removed; remaining values were median-imputed, zero-variance features removed and features with |Spearman r| ≥0.90 pruned. Seventeen XGBoost[100] hyperparameter configurations and four feature-set sizes (10, 20, 30 and 50), ranked by univariable F statistic, were evaluated using inner-fold out-of-fold AUC. Features were centred and scaled using training-fold estimates. Hyperparameters and feature-set size were selected within each outer training set. Platt calibration used out-of-fold predictions generated within the outer training set. Each patient received a single held-out prediction from the corresponding outer-fold model. Elastic-net logistic regression and linear support-vector machines were evaluated using the same outer folds.

***Performance.*** For external validation, Platt calibration was fitted using four-fold out-of-fold predictions from the full FUSCC cohort, and the preprocessing, feature selection and XGBoost model were refitted on all 282 patients. The final calibrated model and the Youden cutoff of 0.6088, derived from the five outer-fold held-out predictions, were applied unchanged across all four external centres. ROC AUCs against reviewed MDP labels used 95% confidence intervals from 5,000 label-stratified patient bootstrap samples of fixed predictions. Classification performance at the locked threshold was assessed using confusion matrices, sensitivity, specificity, positive predictive value and negative predictive value (**Supplementary Table 29**). External calibration was assessed using calibration curves, calibration intercepts and slopes, and Brier scores (**Supplementary Fig. 6**). TreeSHAP estimated feature contributions using each patient's held-out fold model.[101]

***Survival analyses.*** Kaplan–Meier estimates, log-rank tests and Cox models compared predicted MDP groups at the locked threshold. Adjusted models included age per ten years, FIGO stage, R0 and log2-transformed CA125 and HE4. Internally, laboratory values reported beyond measurement limits were set to the stated limits. External analyses used centre-reported follow-up times and event indicators. Event counts for each cohort are provided in **Supplementary Table 27**. Internal and pooled external Cox models used Efron ties and Wald confidence intervals; pooled external adjusted models stratified baseline hazards by centre. Given the limited number of events within individual external centres, centre-specific unadjusted and adjusted analyses used Firth Cox regression (coxphf v1.13.4; Breslow ties), with 95% profile penalised-likelihood confidence intervals and penalised likelihood-ratio tests.[102]

**Statistical analysis**

Tests were two-sided unless specified otherwise. Clinical, malignant-cell, ecosystem and histological comparisons used patient-level data; specimen-, AOI- and spot-level analyses are identified above. Benjamini–Hochberg correction applied within the stated testing families; q denotes adjusted P values. Nominal P values are explicitly identified for exploratory spatial and tissue comparisons.

**Data availability**

The sequencing data generated in this study are being deposited in the Genome Sequence Archive for Human (GSA-Human) under submission ID subHRA032831. The accession number will be provided upon completion of repository processing. The curated and integrated HGSOC scRNA-seq atlas can be downloaded from Zenodo (https://zenodo.org/records/23737155).[103] Other publicly available datasets analysed in this study are described in the relevant Methods sections.

**Code availability**

Code implementing the computational methods developed in this study is available at https://github.com/pigudog/ovarian-cancer-dissemination. Additional scripts for data processing, statistical analyses and reproduction of the main figures are available at https://github.com/pigudog/hgsoc-miliary-atlas.

**Methods References**

**Acknowledgements**

We thank the patients and their families for their participation and the clinical teams for their support. This work was supported by the National Natural Science Foundation of China (82272898, 82203723, 82271940 and 82471932), the Three-Year Action Plan to Promote Clinical Skills and Clinical Innovation Capacity of Municipal Hospitals by the Shanghai Shenkang Hospital Development Center (SHDC2020CR5003-001, SHDC2025CCS006 and SHDC2026CCS006), the Science and Technology Commission of Shanghai Municipality (21ZR1415000), the CSCO-CYH Oncology Research Fund (Y-Young2024-0297), and the Autonomous Region Key Research and Development Program (2024B03037-1).

**Author contributions**

Y.Y.Z., Z.L.W. and S.W. conceived and designed the study. Y.Y.Z., Z.L.W., S.W. and C.H. developed the methodology, with S.W. leading the development of the AI framework and computational algorithms. S.N. contributed to the refinement of anatomical definitions, dissemination phenotyping, validation strategy and clinical interpretation of the findings. Y.Y.Z., Y.Q.C. and Q.H.G. performed statistical analyses and interpreted the results. Z.H.Z., S.L., H.L., J.Z., Y.P.Z., J.D. and J.L. contributed to clinical data collection, curation and analysis. Y.Z., X.Z., X.X.Z. and X.F.C. contributed multicentre imaging data and external validation. Z.H.Z., Z.J.L. and H.M.L. contributed to imaging analysis and radiomics modelling. X.C.W., Y.Q.Z., H.Y.Z. and W.J. provided pathological specimens, experimental materials and research resources. Z.J.L., Z.F., H.W. and X.Z.J. contributed to data processing, annotation and organisation. Z.L.W. and S.W. developed data visualisations and prepared figures. Y.Y.Z. and Z.L.W. wrote the original draft. S.N., C.H., X.H.W., S.W., H.M.L. and Q.H.G. reviewed and edited the manuscript. C.H., S.W. and X.H.W. supervised the research. H.M.L., X.H.W. and Q.H.G. coordinated the project and managed research resources. H.M.L., X.H.W. and Q.H.G. acquired funding. All authors reviewed and approved the final manuscript.

## Competing interests

The authors declare no competing interests.